\documentclass[twocolumn]{aastex63}

\usepackage{xcolor}
\graphicspath{{./}{figures/}}

\received{January 9, 2023}
\accepted{April 29, 2023}

\shorttitle{}
\shortauthors{Wijsen et al.}

\newcommand{\vect}[1]{\mathbf{#1}}
\renewcommand{\vec}{\vect}

\newcommand{\RevOne}[1]{{\textcolor{black}{#1}}}
\newcommand{\RevTwo}[1]{{\textcolor{black}{#1}}}
\newcommand{\RevThree}[1]{{\textcolor{black}{#1}}}

\begin{document}


\title{The effect of the ambient solar wind medium on a CME-driven shock and the associated \RevOne{gradual} solar energetic particle event}

\correspondingauthor{Nicolas Wijsen}
\email{nicolas.p.wijsen@nasa.gov}

\author[0000-0001-6344-6956]{Nicolas Wijsen}
\affiliation{Heliophysics Science Division, NASA Goddard Space Flight Center, Greenbelt, MD 20771, USA}
\affiliation{Department of Astronomy, University of Maryland College Park, MD 20742, USA}

\author[0000-0002-3176-8704]{David Lario}
\affiliation{Heliophysics Science Division, NASA Goddard Space Flight Center, Greenbelt, MD 20771, USA}

\author[0000-0003-0277-3253]{Beatriz S{\'a}nchez-Cano}
\affiliation{School of Physics and Astronomy, University of Leicester, Leicester, LE1 7RH, UK}

\author[0000-0002-0606-7172]{Immanuel C. Jebaraj}
\affiliation{Centre for mathematical Plasma-Astrophysics, KU Leuven, 3001 Leuven, Belgium}

\author[0000-0003-3903-4649]{Nina Dresing}
\affiliation{Department of Physics and Astronomy, University of Turku, FI-20500 Turku, Finland}

\author[0000-0002-3855-3634]{Ian G. Richardson}
\affiliation{Heliophysics Science Division, NASA Goddard Space Flight Center, Greenbelt, MD 20771, USA}
\affiliation{Department of Astronomy, University of Maryland College Park, MD 20742, USA}

\author[0000-0003-1539-7832]{Angels Aran}
\affiliation{Dep. F\'{\i}sica Qu\`antica i Astrof\'{\i}sica. Institut de Ci\`encies del Cosmos (ICCUB), Universitat de Barcelona (UB-IEEC), 08028 Barcelona, Spain}

\author[0000-0001-6589-4509]{Athanasios Kouloumvakos}
\affiliation{The Johns Hopkins University Applied Physics Laboratory, Laurel, MD 20723, USA}
\author[0000-0002-9829-3811]{Zheyi Ding}
\affiliation{Centre for mathematical Plasma-Astrophysics, KU Leuven, 3001 Leuven, Belgium}
\author[0000-0002-3746-9246]{Antonio Niemela}
\affiliation{Centre for mathematical Plasma-Astrophysics, KU Leuven, 3001 Leuven, Belgium}
\affiliation{Solar-Terrestrial Centre of Excellence—SIDC, Royal Observatory of Belgium, 1180 Brussels, Belgium}

\author[0000-0001-6590-3479]{Erika Palmerio}
\affiliation{Predictive Science Inc., San Diego, CA 92121, USA}
\author[0000-0003-1758-6194]{Fernando Carcaboso}
\affiliation{Heliophysics Science Division, NASA Goddard Space Flight Center, Greenbelt, MD 20771, USA}
\affiliation{Physics Department, The Catholic University of America, Washington, DC 20064, USA}

\author[0000-0002-3298-2067]{Rami Vainio}
\affiliation{Department of Physics and Astronomy, University of Turku, FI-20500 Turku, Finland}

\author[0000-0001-9325-6758]{Alexandr Afanasiev}
\affiliation{Department of Physics and Astronomy, University of Turku, FI-20500 Turku, Finland}
\author[0000-0002-5712-9396]{Marco Pinto}
\affiliation{European Space Research and Technology Centre, European Space Agency, 2201 AZ Noordwijk, Netherlands}

\author[0000-0002-6176-4077]{Daniel Pacheco}
\affiliation{Institut für Experimentelle und Angewandte Physik, Christian-Albrechts-Universität zu Kiel, 24118 Kiel, Germany}

\author[0000-0002-1743-0651]{Stefaan Poedts}
\affiliation{Centre for mathematical Plasma-Astrophysics, KU Leuven, 3001 Leuven, Belgium}
\affiliation{Uniwersytet Marii Curie-Skłodowskiej, 20-400 Lublin, Lubelskie, Poland}

\author[0000-0001-7894-8246]{Daniel Heyner}
\affiliation{Institut für Geophysik und extraterrestrische Physik, Technische Universität Braunschweig,  38106 Braunschweig, Germany}

\begin{abstract}
We present simulation results of a gradual solar energetic particle (SEP) event detected on 2021  October 9 by multiple spacecraft, including BepiColombo (Bepi) and near-Earth spacecraft such as the Advanced Composition Explorer (ACE). A peculiarity of this event is that the presence of a high speed stream (HSS) affected the \RevOne{low-energy ion component ($\lesssim 5$~MeV) of the gradual SEP event} at both Bepi and ACE, despite the HSS having only a modest solar wind speed increase. 
Using the EUHFORIA (European Heliospheric FORecasting Information Asset) magnetohydrodynamic model, we replicate the solar wind during the event and the coronal mass ejection (CME) that generated it. We then combine these results with the energetic particle transport model PARADISE (PArticle Radiation Asset Directed at Interplanetary Space Exploration).
We find that the structure of the CME-driven shock was affected by the non-uniform solar wind, especially near the HSS, resulting in a shock wavefront with strong variations in its properties such as its compression ratio and obliquity. By scaling the emission of energetic particles from the shock to the solar wind compression at the shock, an excellent match between the PARADISE simulation and in-situ measurements of $\lesssim 5$~MeV ions is obtained. Our modelling shows that the intricate intensity variations observed at both ACE and Bepi were influenced by the non-uniform emission of \RevOne{energetic particles} from the deformed shock wave and demonstrates the influence of even modest background solar wind structures on the development of SEP events. 
\end{abstract}

\section{Introduction}
When solar energetic particles (SEPs) escape from their acceleration site they propagate through the solar wind and may eventually be detected in-situ by particle detectors on board spacecraft.
Since SEPs tend to follow the interplanetary magnetic field (IMF) configuration, a prompt onset in the SEP intensity--time profiles indicates a good magnetic connection between the spacecraft and the SEP source, which is usually located west of the spacecraft due to the nominal spiral shape of the IMF lines \citep[e.g,][]{cane88}.
However, the occurrence of SEP events in a disturbed interplanetary medium renders the association between the observed intensity enhancements and the SEP source more difficult.

Although most SEP models assume a nominal Parker spiral magnetic field \citep[e.g.,][]{whitman22}, the presence of intervening solar wind structures may modify the configuration of the magnetic field. 
Such structures include interplanetary coronal mass ejections (ICMEs), stream interaction regions (SIRs), and folds in the tilted heliospheric current sheet (HCS), among others \citep[e.g.,][]{kallenrode_cliver01, bieber_02, masson12,  lario14a,palmerio21,smith01,richardson04,richardson18,richardson96}. 
Apart from modifying the IMF geometry, the magnetic field distortions caused by these structures can affect the transport of SEPs by, for example, creating magnetic reflection regions and altering the nominal path length followed by the particles \citep[e.g.,][]{bieber_02, wijsen20}. 
Interplanetary structures can also have a direct impact on the turbulence in the medium through which the SEPs propagate \citep[e.g.,][]{smith_11, kilpua20}. 
Moreover, the characteristics of the shock waves propelled by fast CMEs, \RevOne{which often continuously accelerate energetic particles as they propagate away from the Sun}, can be altered by the intervening structures encountered upstream.\citep[e.g.,][]{odstrcil_pizzo_99, case08, richardson10}.
The interaction between a shock and these aforementioned intervening structures may alter the shock properties, which can affect the particle acceleration efficiency and, consequently, modify the \RevOne{time profiles of the energetic particle intensities} measured by spacecraft. 
 \RevOne{These energetic particle enhancements are known as gradual SEP events, and include all energetic particles accelerated by the shock driven by a CME as it travels through interplanetary space \citep[e.g.,][]{reames99,desai16}. The non-uniformity of the solar wind medium can create a variety of SEP intensity time profiles, each with unique characteristics to understand.}

An intricate SEP event where such intervening structures played a fundamental role in the in-situ energetic particle  measurements  occurred on 2021 October 9. The observational characteristics of this event were studied in detail by \citet{lario22}. 
The SEPs associated with this eruptive event, \RevOne{which included protons up to ${\sim}$100 MeV}, were detected by various spacecraft, including Solar Orbiter \citep[][]{muller20}, Parker Solar Probe \citep[][]{fox16}, BepiColombo \citep[Bepi;][]{benkhoff21}, the Solar Terrestrial Relations Observatory Ahead \citep[STEREO-A;][]{kaiser08}, the
Advanced Composition Explorer \citep[ACE;][]{stone98}, and the Wind spacecraft \citep[][and references therein]{wilson21}. 
These spacecraft were located in a narrow range of longitudes extending from $48^\circ$ east to $2^\circ$ west of Earth and at heliocentric distances from 0.33 to 1~au. The solar eruptive event producing the SEPs included a CME and an 
M1.6-class  solar flare in soft X-rays (2B in H$\alpha$) that originated from the NOAA Active Region  12882, located at N17$^\circ$E09$^\circ$. Hence, the spacecraft were well-located close to the longitude of the solar event. The electron event and the associated radio emission were studied in detail by \citet{Jebaraj22}.

One of the main conclusions of \citet{lario22} was that the intensity--time profiles of the energetic ions measured at Bepi (located at 0.33~au) as well as ACE and Wind (both at L1) were strongly affected by an intervening high speed stream (HSS), despite the fact that the HSS was characterised by only a modest gradual increase in the solar wind speed from $\sim$300 to 410~km s$^{-1}$ at the Sun-Earth  Lagrangian point L1.  
\RevThree{
This modest gradual increase in the solar wind speed and the absence of shock waves bounding the SIR generated by the HSS suggest that the SIR alone did not significantly contribute to the acceleration of the observed energetic particles. This is supported by the fact that STEREO-A (at $30^\circ$ East of Earth) detected only a marginal increase (less than a factor of two) in suprathermal ($\le 350$~keV) proton intensities when the SIR crossed the spacecraft on October 7, before the SEP event occurred. Furthermore, intensities observed near Earth during the SEP event modelled here were significantly higher than those typically seen during pure SIR events \citep[e.g.,][]{lee10}, indicating that the event was indeed an SEP and not an SIR event.
}

During the onset of the SEP event, Bepi and the spacecraft at L1, separated by 0.67~au, were approximately radially aligned and thus on different nominal spiral magnetic field lines.
Nevertheless, both Bepi and the L1 spacecraft measured similarly shaped intensity--time profiles  for ion energies $\lesssim$5 MeV. 
\citet{lario22} suggested that the SEP intensity--time profiles observed at Bepi and ACE or Wind could be the result of particles being confined between the SIR driven by the HSS and the CME that generated the SEP event. 
\RevOne{In this work, we build on the study of \citet{lario22} by further exploring the impact of the intervening HSS on the low-energy component ($\lesssim$5 MeV) of the gradual SEP event observed by Bepi and near-Earth spacecraft. To do so, we use a modelling approach that combines the particle transport code `PArticle Radiation Asset Directed at Interplanetary Space Exploration'  \citep[PARADISE;][]{wijsen19a,wijsenPHD20} with the data-driven solar wind and CME propagation model `EUropean Heliospheric FORcasting Information Asset' \citep[EUHFORIA;][]{pomoell18,Poedts2020}.
Since the inner boundary of EUHFORIA is located at 21.5 solar radii, we do not model particles accelerated early in the SEP event, including high energy protons which are presumably mostly accelerated when the shock is still close to the Sun \citep[e.g.,][]{gopalswamy05,reames09a,reames09b}. Rather, we focus on protons with energies $\lesssim 5$~MeV and assume that  most of these protons observed by near-Earth spacecraft and Bepi are produced by the shock at larger heliocentric distances, when these spacecraft are magnetically connected with the shock front \citep{lario22}.
} 

The EUHFORIA simulation indicates that the large-scale structure of the shock wave driven by the CME as well as the CME itself were strongly deformed due to the non-uniform upstream solar wind conditions.
 This in turn led to a shock wave with strongly varying properties, such as its obliquity and compression ratio. 
By assuming that the SEP production is proportional to the compression of the shock wave, a good match is found between the observations and the simulation \RevOne{for $\lesssim$5~MeV protons}. 
This suggests that acceleration processes at the shock \RevOne{driven by the ICME}
may have played a predominant role in shaping the intensity--time profiles of the \RevOne{low-energy component} of the gradual SEP event at Bepi and near Earth. 
It also demonstrates the importance of having realistic models for the solar wind and CME-driven shocks when trying to understand SEP events, as noted in previous studies \citep[e.g.,][]{lario17,Kouloumvakos2022}. 

The paper is structured as follows. 
In Section~\ref{sec:euhforia}, we present the EUHFORIA simulation of the ambient solar wind and the CME. 
In particular, we emphasise how the modelled CME-driven shock was strongly deformed.
Section~\ref{sec:paradise} provides the results of the PARADISE simulation, and a comparison between the observations and the simulations is presented. 
Section~\ref{sec:summary} summarises the main results of the present work.

\section{The solar wind and shock deformation}\label{sec:euhforia}

\begin{figure*}
    \centering
\includegraphics[width=0.99\textwidth]{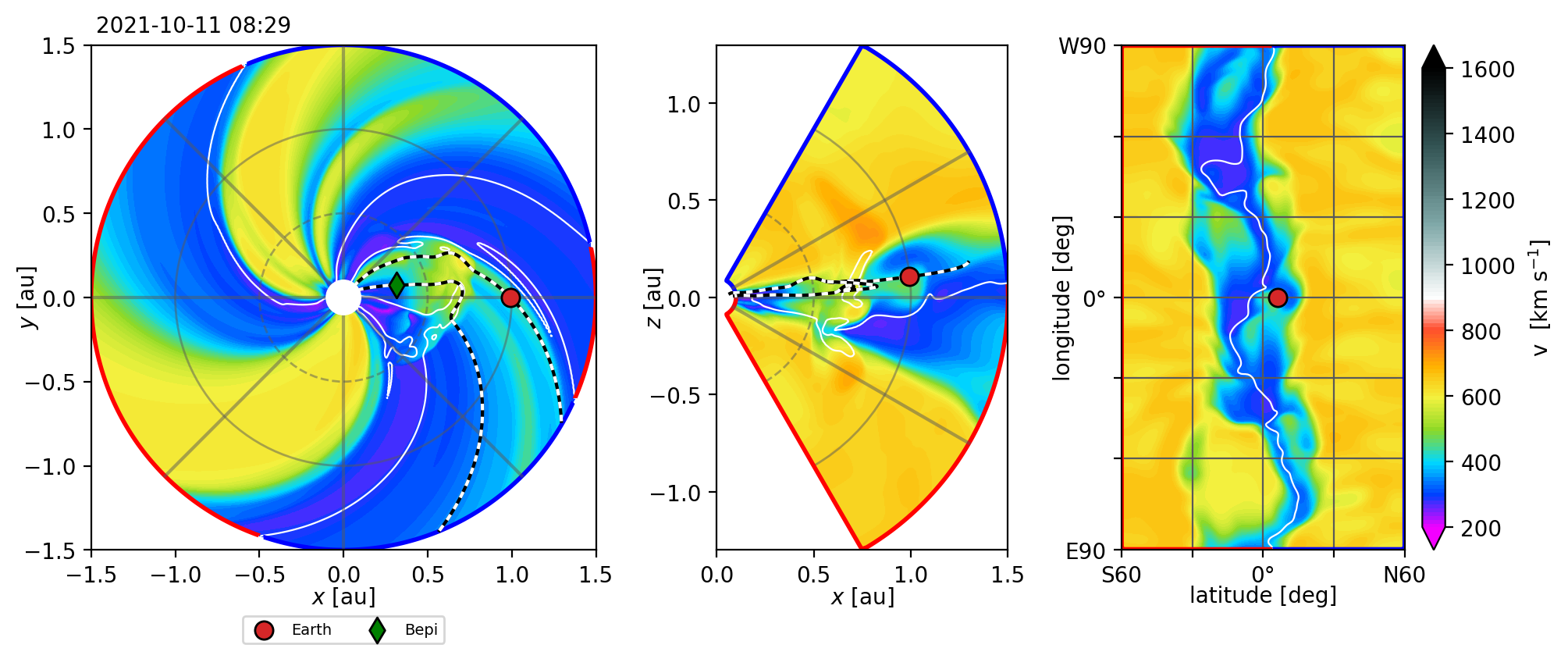}

    \caption{
    Snapshot (at 2021 October 11, 08:29~UT) of the solar wind speed modelled with EUHFORIA. From left to right, the panels show the solar equatorial plane, the meridional slice containing Earth (red circle), and a slice at $r=1$~au. The black--white dashed lines are projected magnetic field lines connecting to Earth and to Bepi (green diamond) and the white thin lines indicate the heliospheric current sheet. \RevOne{An animated version of this figure is available online, showing the propagation of the CME through the solar wind.}
    }
    \label{fig:solar_wind_sim_v}
\end{figure*}

\begin{figure*}
    \centering
\includegraphics[width=0.99\textwidth]{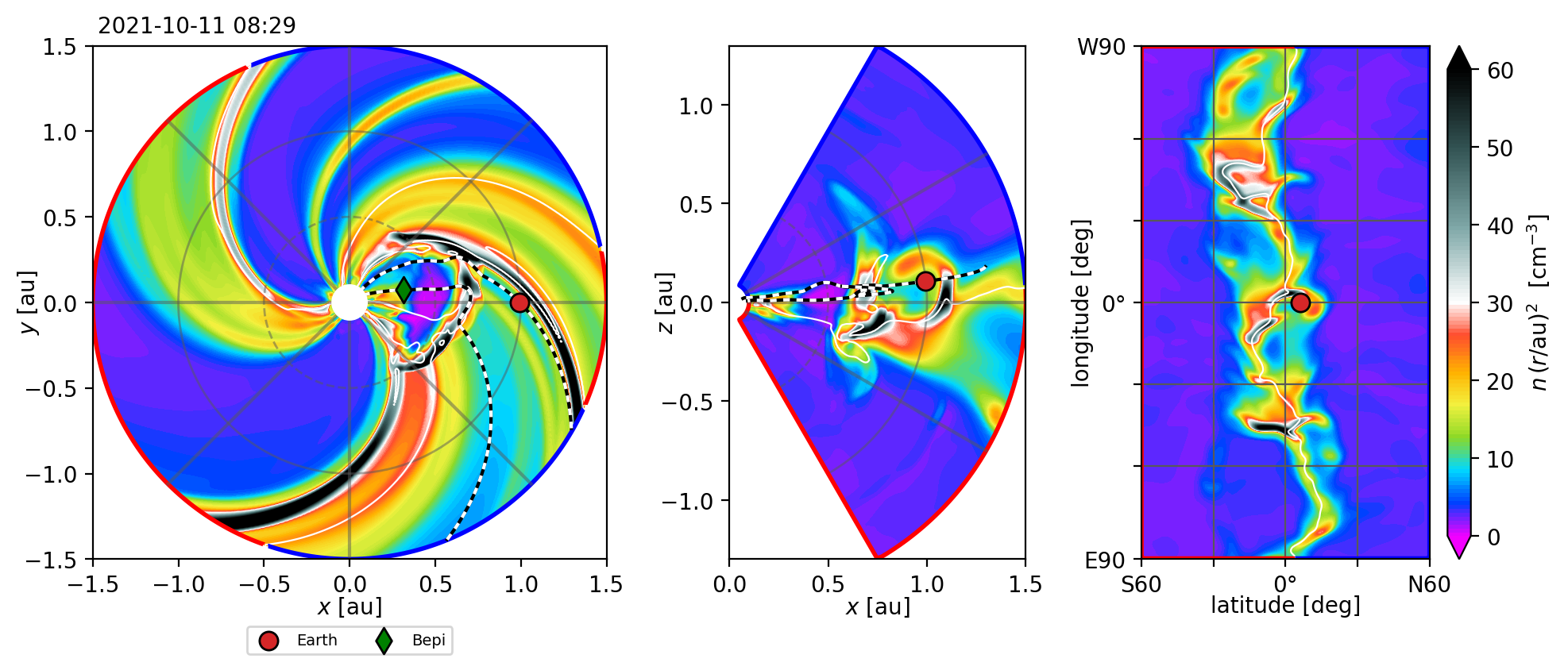}
    \caption{
    Same as Figure~\ref{fig:solar_wind_sim_v}, but showing the scaled number density. \RevOne{An animated version of this figure is available online, showing the propagation of the CME through the solar wind.}
    }
    \label{fig:solar_wind_sim_n}
\end{figure*}

\begin{table}
\caption{\label{tab:cone}Input parameters of the cone CME model in the EUHFORIA simulation.}
\centering
\begin{tabular}{lccc}
\hline\hline
Parameter &Value\\
\hline
Insertion time           &  2021 Oct 9 at 10:30 UT\\
Insertion speed           &  650 km s$^{-1}$\\
Insertion longitude (HEEQ)      & 0$^\circ$  \\
Insertion latitude (HEEQ) &  $6^\circ$ \\
Half width          &  45$^\circ$\\
Density     &    $10^{-18}$ kg m$^{-3}$\\
Temperature      &  $8$ $\times$ $10^5$ K\\
\hline
\end{tabular}
\end{table}

Figure~\ref{fig:solar_wind_sim_v} shows a snapshot of the solar wind speed as obtained from the EUHFORIA simulation in, from left to right, the solar equatorial plane, a meridional slice containing Earth (indicated by a red circle), and a longitudinal versus latitudinal surface at $r = 1$~au. 
\RevOne{Figure~\ref{fig:solar_wind_sim_n} shows the same snapshot as Figure~\ref{fig:solar_wind_sim_v}, but for the scaled number density instead of the solar wind speed.}
The synoptic magnetogram and the solar wind speed map at 0.1 au used as input for the EUHFORIA simulation of the background solar wind is shown in Figure~3 of \citet{lario22}.
In \RevOne{Figure~\ref{fig:solar_wind_sim_v}}, the intervening HSS can be seen arriving at Earth at the time of this snapshot (2021 October 11, 08:29~UT).  
The density enhancement that has just passed Earth in the left and middle panel \RevOne{of Figure~\ref{fig:solar_wind_sim_n}} is the SIR produced by this HSS.
In addition, both figures show the CME propagating from the Sun toward Earth at ${\sim}0.6$~au (where the speed and density show a sudden increase).
The CME is simulated using EUHFORIA's cone model, which consists of a hydrodynamic cloud of plasma of elevated density and temperature that is inserted into the solar wind with a constant speed and angular width \citep[e.g.,][]{scolini18}. 
The insertion parameters of the cone CME are presented in Table~\ref{tab:cone}. 
For the CME's density and temperature, EUHFORIA's default parameters \citep[see][]{pomoell18} were used and the kinematic insertion parameters were chosen by slightly adjusting the shock fitting results presented in \citet{lario22}, so that the modelled and observed arrival times of the CME shock at Earth match (see also Figure~\ref{fig:earth}).  

\begin{figure}
    \centering
    \includegraphics[width=0.45\textwidth]{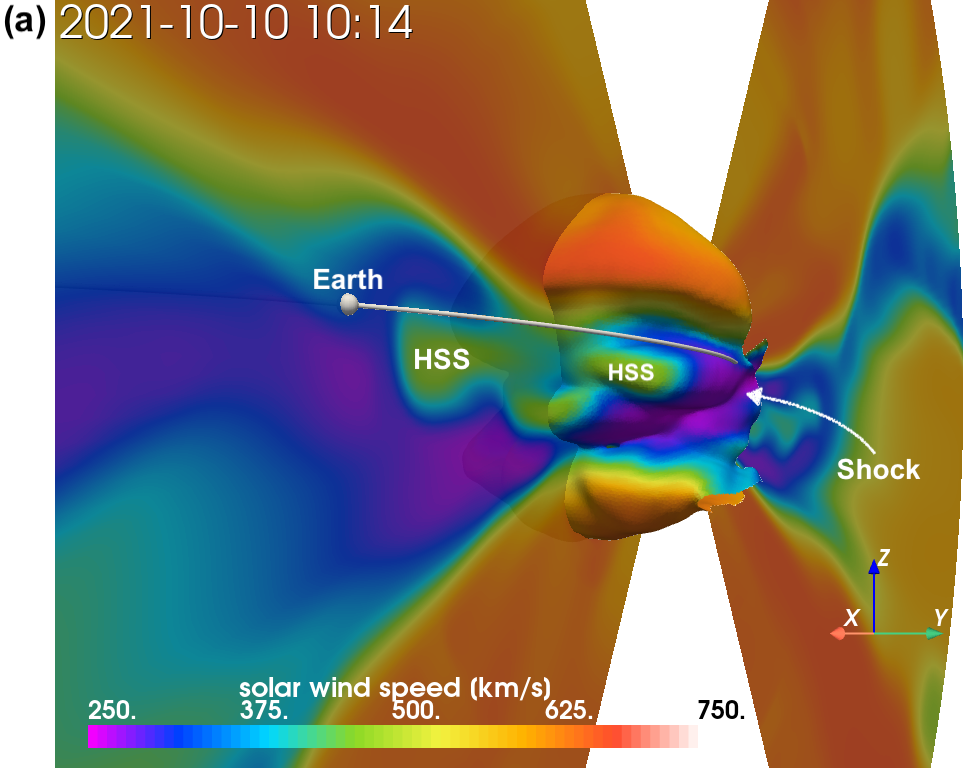}
    \includegraphics[width=0.45\textwidth]{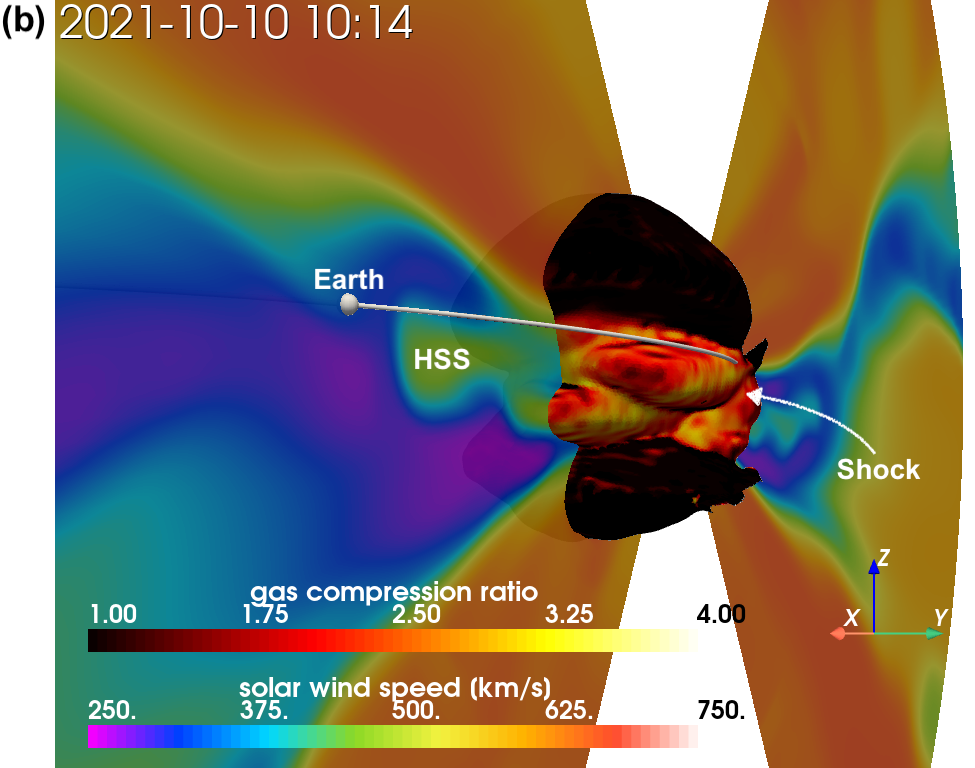}
    \includegraphics[width=0.45\textwidth]{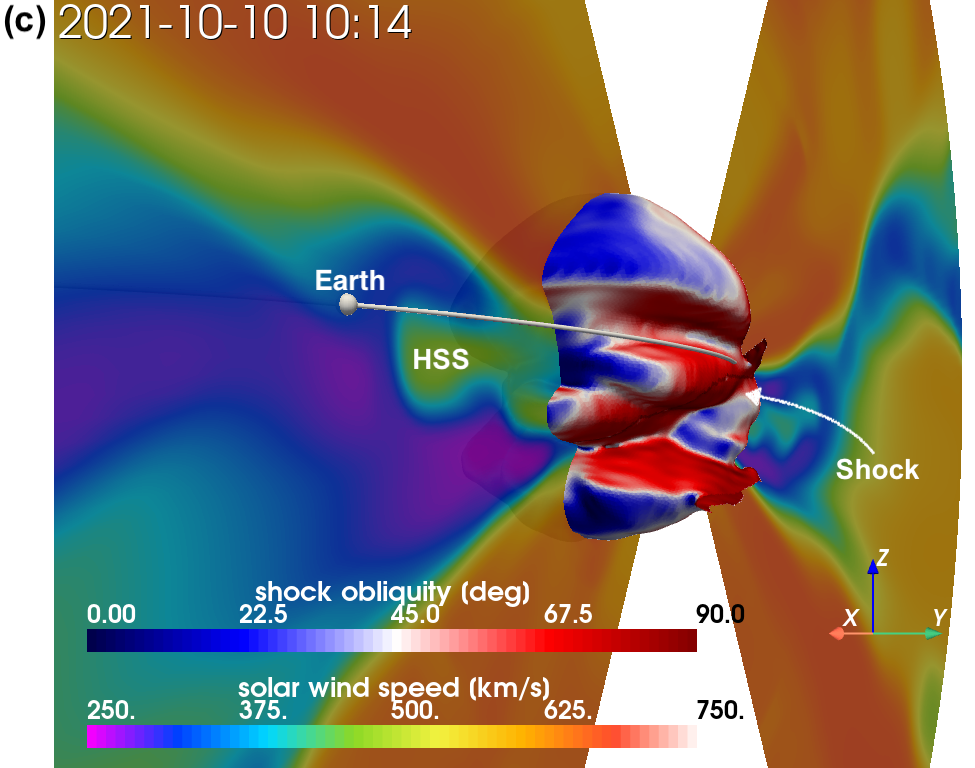}
    \caption{The CME-driven shock surface (occulting the sun) together with a meridional slice colour-coded according to the solar wind speed. The magnetic field line connecting Earth with the shock surface is shown in white. In panel (a), the shock surface is colour-coded according to the upstream solar wind speed, in panel (b) according to the gas compression ratio $r_g$, and in panel (c) according to the shock obliquity $\theta_{B_n}$.
    The panels correspond to the same time (${\sim}$23.75 hours after the insertion of the CME) and capture the shock when its nose was at ${\sim}0.45$~au. }
    \label{fig:shock}
\end{figure}

What is evident from \RevOne{Figures~\ref{fig:solar_wind_sim_v} and~\ref{fig:solar_wind_sim_n}}  is that the modelled CME (and its shock wave) was distorted during its propagation through the inner heliosphere. 
This distortion can be attributed to the varying solar wind conditions upstream of the CME \citep[e.g.,][]{savani10,owens17}. 
That is, the portions of the CME propagating through the HSS are less decelerated than the portions propagating through the slow solar wind trailing and preceding the HSS. 
This is because the larger density and the slower speed of the slow solar wind lead to a drag force acting on the CME that is greater than in the fast solar wind.

To further illustrate the deformation of the CME, we show in Figure~\ref{fig:shock} a meridional slice showing the ambient solar wind speed together with the shock wavefront of the CME, which was obtained using EUHFORIA's shock tracer \citep[see][for details]{wijsen22}. 
The panels show the western flank of the CME, which is the flank to which Earth is magnetically connected when the CME propagates away from the Sun.
The magnetic field line connecting Earth (white circle) with the shock is shown as a white line.
The figure illustrates how the non-uniform upstream solar wind conditions, in addition to deforming the shock surface, also cause the shock to have strongly varying properties. In panel~(a), the shock surface is colour coded according to the upstream solar wind speed, with the HSS affecting the shock surface. In panel~(b) of Figure~\ref{fig:shock}, the shock surface is colour-coded according to its gas compression ratio $r_g$, that is, the ratio between the downstream and upstream plasma density (black-orange colour bar).  
Where the surface is coloured black,  the CME's propagation speed is lower than the local upstream fast magnetosonic speed and hence in those regions the depicted surface is not a shock wave. 
This occurs mostly in the fast wind originating from the highest northern and southern latitudes. 
Similarly, it can be seen that where the shock is propagating through the HSS, the  compression ratio is $r_g\lesssim 2$ (black-reddish), whereas in the preceding and trailing slow wind, the compression ratio is $r_g\gtrsim 2.5$ (yellowish). 

In panel~(c) of Figure~\ref{fig:shock}, the shock is colour-coded according to the shock obliquity $\theta_{B_n}$, which is defined as the angle between the upstream magnetic field and the shock normal (blue--red colour bar).  
A classic cartoon representation  assumed for
a CME propagating through a uniform solar wind shows that the shock obliquity  smoothly varies from a quasi-perpendicular geometry on the west flank of the CME to a quasi-parallel geometry on the east flank of the CME (e.g., Figure~11 in \citet{sarris84} or Figure~6 in \citet{zank06}). 
This is because it is assumed that
(1) the IMF  has the same spiral shape everywhere, and 
(2) the CME does not show a markedly deformation, apart from some flattening, provided that there are no large inhomogeneities present in the CME itself. 
Both these properties break down once the ambient medium through which the CME propagates is no longer uniform, in which case the shock obliquity is less well behaved. This is because the deformation of the shock wave changes the curvature of the shock front locally, which in turn modifies the shock's local obliquity. 
This is illustrated in panel~(c) of Figure~\ref{fig:shock}, where it can be seen that the west flank of the shock contains both quasi-parallel and quasi-perpendicular regions because of its deformation. Likewise, the eastern flank also contains a mix of both quasi-perpendicular and quasi-parallel geometries (not shown).

\section{Energetic particles at Earth and Bepi}\label{sec:paradise}

\subsection{PARADISE set-up}

Next, we use PARADISE \citep{wijsenPHD20} to model the temporal and spatial evolution of an energetic proton population propagating through the EUHFORIA solar wind configuration presented in the previous section. 
PARADISE does this by solving the 5-dimensional focused transport equation  \citep[FTE; see e.g.,][for a recent review]{vandenberg20} \RevTwo{by integrating an equivalent set of It\^o-stochastic equations forward in time. The resulting pseudo-particles are sampled on a spherical mesh of radial resolution $dr=0.02$~au and angular resolution $d\varphi=d\vartheta=1^\circ$, where $\varphi$ and $\vartheta$ denote the azimuthal and latitudinal coordinates, respectively.
The solution obtained by the model is the directional differential intensity.}
\RevTwo{The FTE solved by PARADISE takes} into account the effects of solar wind turbulence by including a diffusion process in the particles' pitch-angle coordinate and a spatial diffusion process perpendicular to the average IMF. 
For details on the implementation of PARADISE we refer to \citet{wijsenPHD20}.
In the work presented here, we use standard quasi-linear theory \citep[QLT;][]{jokipii66} to prescribe the pitch-angle diffusion coefficient and assume that the protons propagate with a parallel mean free path $\lambda_\parallel = 0.3 (R/R_0)^{2-q}$~au, where $q=5/3$ is the spectral index of a Kolmogorov turbulence spectrum, $R$ is the particle rigidity, and $R_0 = 43~MV$, which corresponds to the rigidity of a 1~MeV proton. 
The resulting values for $\lambda_\parallel$ fall toward the higher end of the range of parallel mean free paths usually derived from observations \citep[e.g.,][]{bieber94}.
In the PARADISE simulation, particle distributions are also subject to a cross-field diffusion process, characterised by a constant perpendicular mean free path $\lambda_\perp=3\times10^{-4}$~au.  
The assumption that $\lambda_\perp/\lambda_\parallel\sim 10^{-3}$ means that the energetic protons are predominantly propagating along the IMF lines in the simulation. 

\RevTwo{In our simulation, protons with energies between 50~keV and 6~MeV are continuously injected along the entire shock wave. This is done by introducing the following source function in the FTE solved with PARADISE} \citep[e.g,][]{prinsloo19,wijsen22}:
\begin{equation}\label{eq:Q}
    Q(E) = C |\nabla\cdot\vec{V}_{\mathrm{sw}}| (E_0/E)^{3} (r_0/r)^2, 
\end{equation}
\RevTwo{where $C$ is a normalisation factor with the units of differential intensity (cm$^{-2}$ MeV$^{-1}$ s$^{-1}$ sr$^{-1}$), $E$ is the proton energy, and $\vec{V}_{\mathrm{sw}}$ is the solar wind velocity vector. The reference values $E_0$ and $r_0$ are chosen as $88$~keV and $1$~au, respectively}
The divergence of the solar wind velocity vector appearing in Eq.~(\ref{eq:Q}) measures the rate of the compression/expansion of the solar wind \RevThree{in units of s$^{-1}$}.  \RevOne{
Despite prevalent particle acceleration theories, such as diffusive shock acceleration \citep[DSA; e.g.,][]{bell78,drury81} and shock drift acceleration \citep[SDA; e.g.,][and references therein]{ball01}, suggesting a dependence on $\theta_{B_n}$,  our injection rate $Q$ does not consider it explicitly. 
This is partly due to the incomplete understanding of the exact relation between $\theta_{B_n}$ and the ongoing particle acceleration mechanism, and partly because our simulation reproduce the observations well without considering this factor (see Section~\ref{sec:summary} for further discussion). 
Moreover, our simulation does not aim to replicate the exact acceleration mechanism that occurs at the shock wave. Rather, we assume that the particle distributions emitted from the shock, as described in Eq.~(\ref{eq:Q}), are the outcome of the acceleration mechanism happening at the shock.  Although the injected particles may undergo additional acceleration upon interacting with the CME shock wave during the simulation, this extra acceleration is minimal. This is because the particles' mean free path is not significantly reduced near the shock, which prevents them from being efficiently trapped close to the shock. As a result, an efficient DSA process does not take place in the current PARADISE simulation.
}
\begin{figure}
    \centering
    \includegraphics[width=0.49\textwidth]{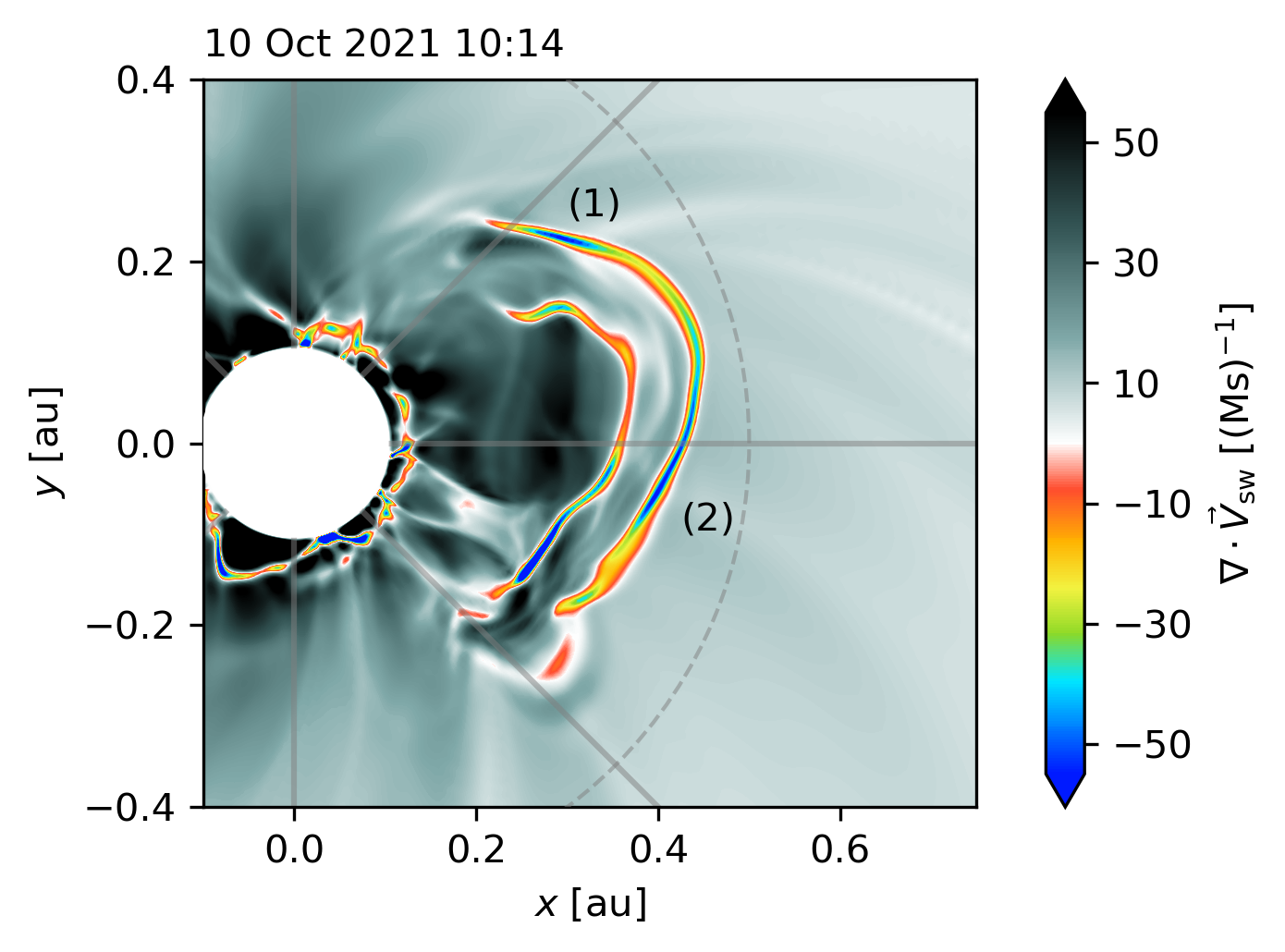}
    \caption{The divergence of the modelled solar wind velocity in the solar equatorial plane, centred on the CME, on 2021 October 10, 10:14~UT when the CME nose was at 0.45~au. \RevThree{The grey dashed semi-circle indicates the heliocentric distance of 0.5~au}. Labels (1) and (2) indicate the regions of highest $-\nabla\cdot\vec{V}_{\rm sw}$. }
    \label{fig:div}
\end{figure}

Figure~\ref{fig:div} shows the divergence of $\vec{V}_{\mathrm{sw}}$ in the solar equatorial plane as modelled by EUHFORIA, 23.75~hours after the insertion of the CME.
Compression and shock waves in the solar wind can be easily identified with $\nabla \cdot \vec{V}_{\mathrm{sw}}$, since these structures are characterised by converging flows and hence $\nabla \cdot \vec{V}_{\mathrm{sw}} < 0$  (indicated by red, yellow and blue colors in the figure). Figure~\ref{fig:div} shows a leading forward shock (approaching the heliocentric distance of ${\sim}$0.45 au) followed by a reverse shock. The formation of these two shocks is a consequence of the cone CME inserted in the EUHOFIRA simulation. 
Figure~\ref{fig:div}  illustrates that along the forward shock front, $\nabla \cdot \vec{V}_{\mathrm{sw}}$  varies similarly to the gas compression ratio $r_g$ shown in Figure~\ref{fig:shock}b. 
That is, the most negative $\nabla\cdot\vec{V}_{\mathrm{sw}}$ values in the shock front can be seen in the regions indicated by (1) and (2) in Figure~\ref{fig:div}, which is where the slow wind is preceding and trailing the HSS, and where $r_g$ is also enhanced.
The correspondence between $\nabla\cdot\vec{V}_{\mathrm{sw}}$ and $r_g$ is not surprising, since both quantities give a measure of the compression of the simulated shock front.
Apart from the CME-driven shock waves, there are some standing  compressive structures close to the inner boundary ($r<0.15$~au), which arise from an imbalance in the total pressure (that is, magnetic and thermal pressure) at the inner boundary of EUHFORIA. 
In the PARADISE simulation, it is assumed for simplicity that $Q$ is zero in these structures and particles are only emitted from the forward shock wave of the CME, \RevOne{from the time when it is injected at the inner boundary}.

The normalisation factor $C$ in Eq.~(\ref{eq:Q}) is fixed by requiring that the simulated \RevOne{particle} intensities at the time of the shock arrival at ACE match the intensity in the $68-115\;$keV energy channel measured by the Low-Energy Magnetic Spectrometer 120 (LEMS120) of the Electron, Proton, and Alpha Monitor \citep[EPAM;][]{gold_98} on board ACE. The energy spectrum between 115~keV and 4.8~MeV of the SEP fluence measured by LEMS120 between the first SEP intensity peak (2021 October 11 at 01:10~UT; see Figure~\ref{fig:earth}a) and the arrival of the CME (2021 October 12 at 01:40 UT) can be fitted by a power law $E^{-\gamma}$, with $\gamma = 2.78$. 
\RevThree{ According to steady-state DSA, the power law index $\gamma = 2.78$ indicates that the compression ratio of the particle scattering centers across the shock, denoted as $r_{\text{sc}}$, is 1.66. Assuming that the scattering centers are frozen into the solar wind plasma, we have $r_{\text{sc}} = r_g$. However, if the scattering centers are, for example, Alfv\'en waves, $r_{\text{sc}}$ may differ from $r_g$ \citep[see e.g.,][]{vainio98, vainio99a}. It is also important to note that the compression ratio varies with time and space (see Figure~\ref{fig:shock}b), meaning that the energy power law observed at the shock crossing likely results from current and past acceleration conditions at the shock wave.}
In Eq.~(\ref{eq:Q}), we inject a slightly softer power law $E^{-3}$ than the one observed when the shock crosses ACE, since SEP transport processes tend to harden the observed energy spectra \citep[e.g,][]{ruffolo95,wijsen20}.
The factor $r^{-2}$ in Eq.~(\ref{eq:Q}) takes into account the expansion of the solar wind into the heliosphere. 

In the following, we focus on \RevOne{the low-energy ($<5\;$MeV) protons} observed by Bepi and Earth only. 
\RevOne{As already commented,} the EUHFORIA modelling domain starts at 0.1~au and therefore does not include particle acceleration and transport in the corona.
\RevOne{This precludes modeling the SEP event as observed at Solar Orbiter, Parker Solar Probe, and STEREO-A adequately because these spacecraft observed a prompt onset \citep[see Figs.~7--9 in][]{lario22}, presumably due to the arrival of SEPs accelerated by the CME-driven shock when the CME was still below 0.1~au. }
We leave the inclusion of the shock at distances $r<0.1$~au for future work.

\subsection{Comparison between the observations and the simulations}\label{sec:sim_vs_obs}

\begin{figure}
    \centering
\includegraphics[width=0.49\textwidth]{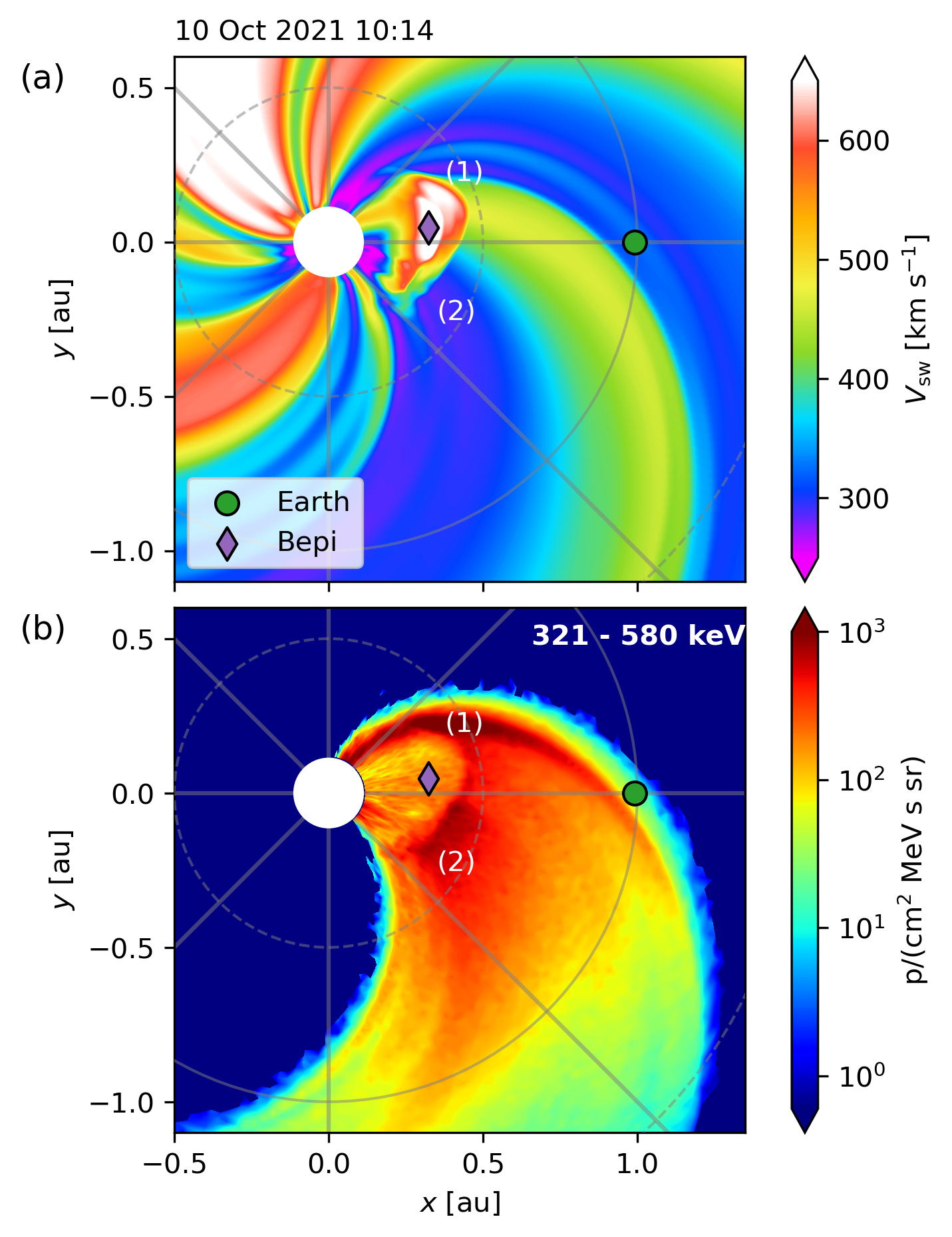}
    \caption{\RevOne{Panel (a) shows a snapshot of the modelled solar wind speed on 2021 October 10, 10:14~UT when the CME nose was at 0.45~au. Panel (b) shows, for the same time, the modelled omnidirectional 321--580 keV protons intensities. Labels (1) and (2) indicate the regions of highest particle intensities (see text for details). An animated version of this figure is available, illustrating the time evolution of the solar wind speed and the omni-directional particle distributions.}}
    \label{fig:v_and_I}
\end{figure}

\RevOne{Figure~\ref{fig:v_and_I} shows the solar wind speed (panel~a) and the simulated omnidirectional $321-580\;$keV proton intensities (panel~b) in a latitudinal slice containing Earth, when the CME nose is at ${\sim}0.45$~au. 
The highest particle intensities are found in the regions indicated by (1) and (2) in the figure, that is, where the shock propagates through the slow wind preceding and trailing the HSS. 
These regions connect magnetically to the most compressive parts of the shock wave where the particle emission is the strongest (regions (1) and (2) indicated in Fig.~\ref{fig:div}).}
In addition, we note that the particle intensities are higher in region (1) than in region (2), despite $|\nabla\cdot\vec{V}_{\mathrm{sw}}|$, and thus $Q$, being slightly larger in region (2). 
This is because region (1) coincides with the SIR that is being driven by the HSS 
and the magnetic field compression inside the SIR causes the particles to propagate along a narrow path, leading to an enhancement of \RevOne{particle} intensities in the elongated dark red band in region (1) in Figure~\ref{fig:v_and_I}b.

\begin{figure*}
    \centering
    \includegraphics[width=0.99\textwidth]{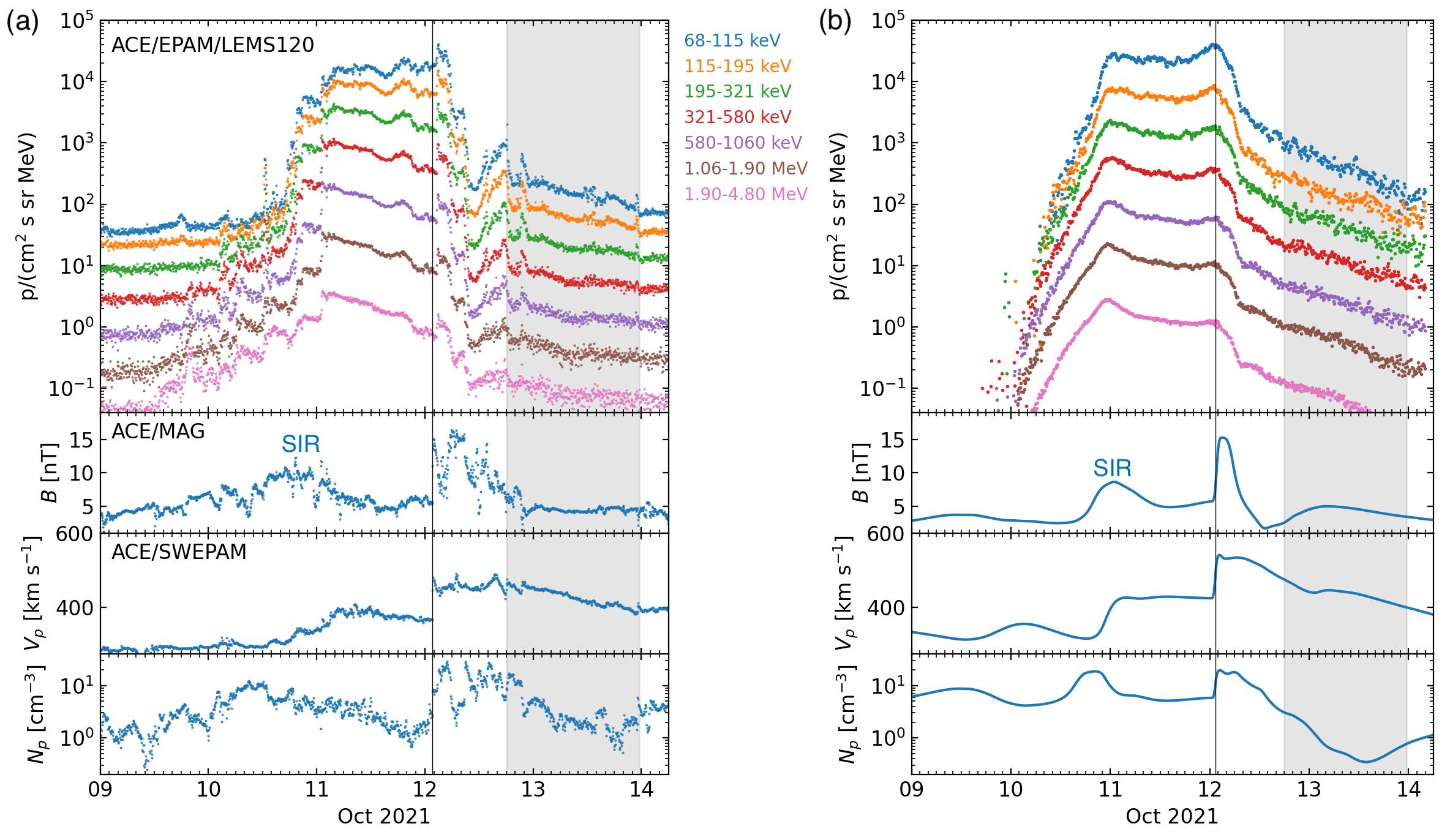}
    \caption{Observed (panel a) and modelled (panel b) SEP intensities and solar wind properties at ACE. From top to bottom, the panels show the SEP intensities for different energy channels spanning 68~keV to 4.8~MeV, the magnetic field magnitude, the solar wind speed, and the proton density. The vertical line indicates the shock arrival time and the grey shaded region indicates the magnetic cloud as identified by \citet{lario22}.  }
    \label{fig:earth}
\end{figure*}

Figure~\ref{fig:earth}a shows the in-situ observations made by ACE near Earth. 
In particular, the figure shows, from top to bottom, the energetic ion intensities from 68~keV to 4.80~MeV  measured by EPAM/LEMS120, the IMF magnitude measured by the Magnetic Field Experiment \citep[MAG;][]{smith98}, and the solar wind proton speed and  density measured by the Solar Wind Electron Proton Alpha Monitor \citep[SWEPAM;][]{mccomas98} on ACE. As discussed by \cite{lario22}, the \RevOne{energetic} ion enhancement at ACE  following the solar event on 2021 October 9 was highly structured and the largest low-energy ion intensities occurred between passage of the SIR indicated in the figure and the shock marked by the solid vertical line. The ion intensities then dropped rapidly in the sheath between the shock and the ejecta of the ICME (grey shaded region).  

Figure~\ref{fig:earth}b shows the omnidirectional proton intensities modelled by PARADISE, together with the magnetic field magnitude, the solar wind speed, and the solar wind proton density modelled by EUHFORIA. 
Although the observed and modelled particle intensities have the same energy ranges, PARADISE considers only protons whereas EPAM/LEMS120 measures ions 
without distinguishing different species. Nevertheless, EPAM/LEMS120 measurements are in principle dominated by protons \citep{gold_98}.  
Figure~\ref{fig:earth} shows that both EUHFORIA and PARADISE simulations successfully reproduce several features of the observed interplanetary medium and SEP event.  
In particular, the modelled intensity--time profiles show a first peak coinciding with the arrival of the SIR. 
This first peak corresponds to the intensity enhancement indicated by number (1) in Figure~\ref{fig:v_and_I}b. 
After the first peak, the modelled intensities decrease slightly for the highest energy channels (${\gtrsim}580$~keV) and remain approximately constant for the lowest energy channels (${\lesssim}321$~keV), in agreement with the observations. 
During this period of approximately flat $\lesssim$321~keV proton intensity--time profiles, the SIR is already beyond ACE, but the spacecraft remains immersed in the HSS and magnetically connected to the SIR at radial distances beyond 1~au. Since the SIR is characterised by a magnetic field enhancement, some outward propagating particles will be mirrored upon reaching the magnetic compression, eventually contributing to the quasi-constant intensities measured by ACE before the arrival of the CME.

The observed sheath region shows several fluctuations in the magnetic field and the proton density that are not captured by our MHD simulation. Both in the observations and in the simulation, the intensities decrease as the sheath passes the observer. In particular, the observed particle intensities decrease in two steps followed by an increase at around ${\sim}$05:00 UT on 2021 October 12 in the sheath prior to the arrival of the ejecta of the ICME (indicated by the grey shaded region). 
The sudden decrease in the observed ion intensities on entry to the ejecta (more prominent at energies ${\lesssim}1$~MeV) is also not reproduced by the simulations that show no change in the rate of the continuing intensity decay at entry into the ICME. This is mostly because the CME was simulated using a simple cone CME model instead of a more sophisticated magnetised CME model \citep[such as the one used in][]{wijsen22}, making access of the particles into the ejecta easier than in the case of a closed magnetic field structure. 

\begin{figure}
    \centering
    \includegraphics[width=0.48\textwidth]{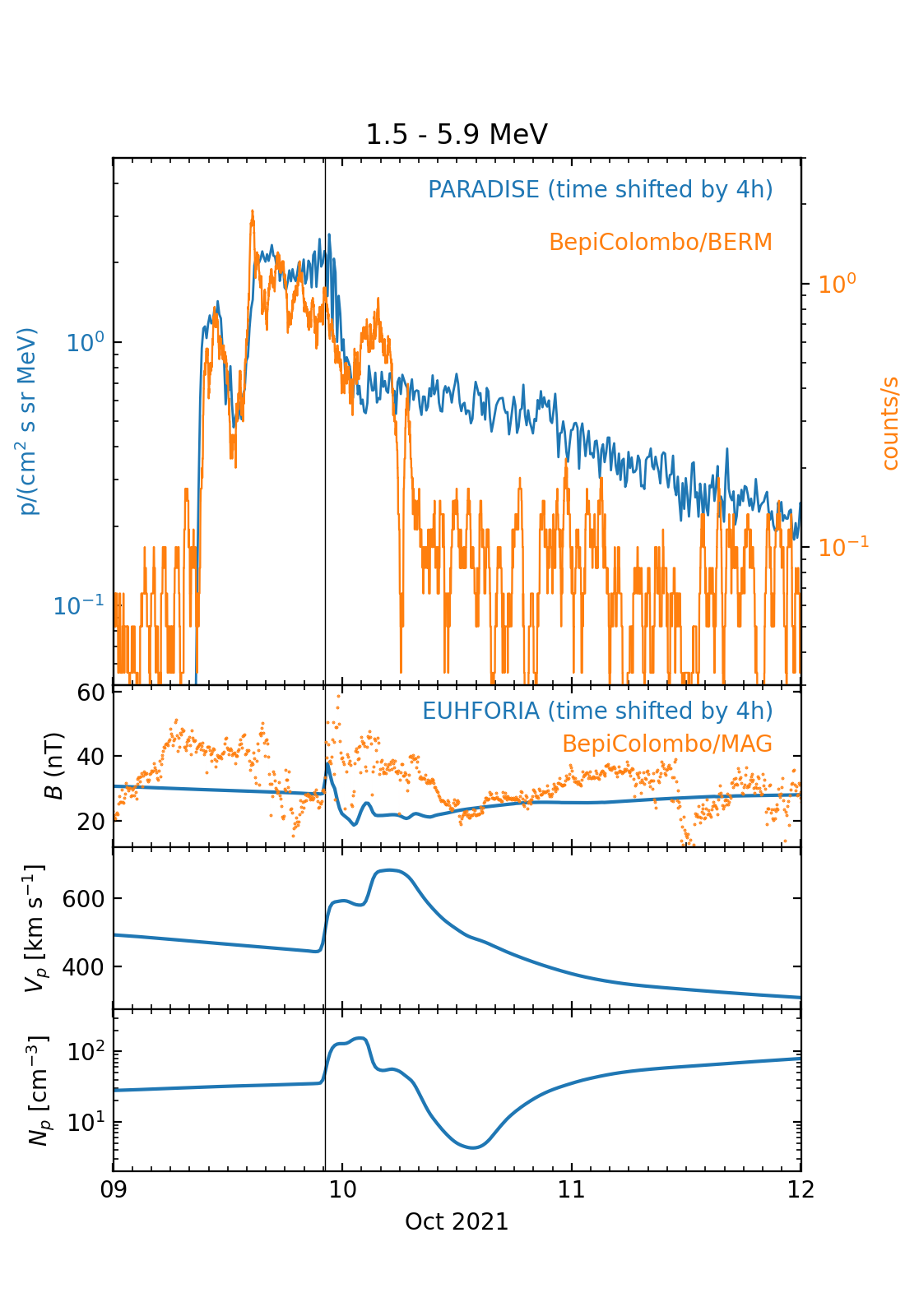}
    \caption{Observed (orange) and modelled (blue) SEP intensity and solar wind properties at Bepi. The PARADISE and EUHFORIA simulation results have been shifted 4 hours back in time to align with the observed shock. The top panel shows the modelled proton intensities together with the BERM ion counts in the energy bin 1.5--5.9 MeV. The second panel shows the observed and modelled magnetic field magnitude. The bottom two panels show the modelled solar wind speed and proton density. }
    \label{fig:bepi}
\end{figure}

Next, we compare the in-situ data and simulation results at Bepi, which was located at 0.33~au from the Sun and $2^\circ$ west of Earth at the onset of the SEP event. 
The top panel of Figure~\ref{fig:bepi} shows the modelled proton intensities (blue line) from PARADISE together with the count rates (orange line) measured in the 1.5--5.9 MeV proton channel of the BepiColombo Radiation Monitor \citep[BERM;][]{pinto22}. 
The second panel of Figure~\ref{fig:bepi} shows the magnetic field magnitude measured by the magnetometer on board Bepi's Mercury Planetary Orbiter \citep[MPO-MAG;][]{heyner21}, together with the modelled magnetic field from EUHFORIA. 
The two bottom panels show the modelled solar wind speed and proton density; unfortunately, no solar wind plasma measurements from Bepi were available.
In order to illustrate the qualitative match between the observed and the simulated SEP intensities, 
the modelled SEP and plasma time series presented in Figure~\ref{fig:bepi} have been shifted backward in time by 4 hours (the reason for this shift is explained below). 
Both the simulated and the BERM intensity--time profiles display a two-peak structure prior to the arrival of the shock (indicated by the vertical solid line).
In the simulation, the first peak is due to the arrival of SEPs accelerated in the interaction between the shock and the developing SIR \RevOne{(region (1) in Figures~\ref{fig:div} and~\ref{fig:v_and_I})} where Bepi was connected to at the onset of the event. The second, longer-duration, increase is due to the approach of the CME-driven shock (i.e., the source \RevOne{of particles}) to Bepi. The profiles do differ following the shock in that the modelled intensity falls rapidly whereas the observations show a further peak followed by an abrupt decrease on entry to the ICME ejecta \citep{lario22}.

We note that the correspondence between the modelled and observed magnetic field is not as good at Bepi as it was at ACE. 
The magnetic field modelled by EUHFORIA is not as structured as the observed one, which may explain why PARADISE misses the intensity peak occurring in the shock sheath. 
Furthermore, the magnetic compression observed at Bepi on 2021 October 9 is not reproduced.  
This magnetic compression was identified by \citet{lario22} as a likely candidate for the developing SIR associated with the HSS observed at ACE, whereas in the simulation, this HSS and the associated magnetic compression passes Bepi earlier, i.e., on 2021 October 8 (this is not shown in  Figure~\ref{fig:bepi}, but easily seen in the movie accompanying Figure~\ref{fig:solar_wind_sim_n}).
Assuming that the magnetic compression seen early on 2021  October 9 at Bepi is indeed the same as the compression seen at ACE late on 2021 October 10, the solar wind must have been travelling at an average speed of $\sim$800 km s$^{-1}$ from Bepi to ACE, which is well above the observed speed at ACE. 
This discrepancy can most likely be attributed to the $4^\circ$ of latitudinal separation between the spacecraft. That is, the coronal hole generating the HSS appeared slanted on the solar disk, extending increasingly westward toward lower latitudes \cite[e.g., H1 in Figure 3b of][]{lario22}. Hence, if the HSS followed a similar spatial structure as its parent coronal hole, two spacecraft located at the same longitude but at different latitudes will see the HSS at different times, with the spacecraft at lower latitudes seeing the HSS earlier (even after adjusting for different helioradii).
Moreover, latitudinal variations inside HSSs and the SIRs they drive are not uncommon, as illustrated by \cite{jian19} who analysed 151 pairs of SIRs seen by STEREO A and B and showed that, even within $5^\circ$ of latitude, the solar wind properties of a single SIR can vary strongly.

As mentioned before, the modelled time series presented in Figure~\ref{fig:bepi} have been shifted back in time by 4 hours. That is, in the original EUHFORIA simulation, the CME shock arrived 4 hours too late at Bepi as compared to the observations. 
A discrepancy of 4 hours between observations and MHD heliospheric models such as EUHFORIA is not uncommon \citep[e.g.,][]{riley18} and, in our case, may be largely due to the mismatch between the observed and simulated HSS at Bepi's latitude.
That is, if the background solar wind is not well captured, the drag on the CME can be over or underestimated making the CME arrive too early or too late. 
An error on the assumed injection parameters of the CME 
and the simplicity of the cone model may also contribute to the 4-hour mismatch between the simulations and observations. 
\RevThree{
Furthermore, we note that, depending on the width and the local expansion of the shock in the lower corona, the interaction between the shock wave and the HSS could have started below the inner boundary of our model, i.e., 0.1 au. Therefore, Bepi may have already established a magnetic connection to the shock before it crossed the 0.1~au boundary, which could contribute to the 4-hour discrepancy between the simulation and observation. Nonetheless, this does not alter the finding that the interaction between the SIR and the CME was most probably responsible for the distinct two-peak particle intensity pattern observed by Bepi.
}

\section{Summary and discussion}\label{sec:summary}

In this study, we present simulation results of the \RevOne{gradual} SEP event observed on 2021 October 9 using the EUHFORIA and PARADISE models and compare these results with the multi-spacecraft observations of this event reported by \cite{lario22}. 
In particular, we focus on understanding the unusual but similar \RevOne{low-energy (${<}5$ MeV) proton} intensity--time profiles measured at Bepi and ACE. According to \cite{lario22}, these observations suggested that the proton intensities were strongly affected by the passage of a corotating high-speed solar wind stream prior to the arrival of the CME shock at both spacecraft. 
An important finding of this work is that the shock driven by the ICME was significantly deformed by the HSS, which led to strongly non-uniform shock conditions that might have changed the efficiency of particle acceleration along the shock front.
In the model, it is  assumed that the \RevOne{particle} emission from the shock is proportional to the  shock strength as measured by $\nabla\cdot\vec{V}_{\mathrm{sw}}$. The excellent agreement between the observed and simulated \RevOne{intensity} time profiles at both Earth and Bepi corroborates the results of the PARADISE simulation. 
The match can be attributed to the significant variations of the modelled shock strength, with the strongest particle emission originating from the regions of the shock propagating through the SIR and the slow wind.


\citet{lario22} hypothesised that the enhanced \RevOne{particle} intensities observed at Earth and Bepi could have been the result of particles confined between the SIR and the approaching CME. 
Although such a transport process might have contributed to shaping the SEP event detected at Bepi and Earth, our modelling suggests that variations of the \RevOne{energetic particle} emission at the CME-driven shock due to the shock interacting with the varying background solar wind, and in particular the SIR, may have played a dominant role.
A similar explanation for another event was proposed by \citet{ding22}, where the authors used the improved Particle Acceleration and Transport in the Heliosphere model \citep[iPATH; see][and references therein]{hu17} to study the SEP event on 2020 November  29, observed by Solar Orbiter, Parker Solar Probe, STEREO-A, and spacecraft near Earth  and Mars \citep{kollhoff21,palmerio22}. 
The CME generating that event also interacted with a high-speed stream, and the authors likewise concluded that this interaction and the resulting deformation of the shock wave played an important role in the variation of
time--intensity profiles measured at different spacecraft. 

Another remarkable feature of the event studied in this work is that the HSS affecting the \RevOne{energetic particle profiles} at Bepi and Earth was in fact quite modest. 
That is, near Earth the solar wind speed only increased from $\sim$300~km s$^{-1}$ to $410$~km s$^{-1}$ during the passage of the HSS. 
The fact that a HSS with such a small speed increase can have a significant effect on the development of the CME shock and the associated SEP event further indicates the necessity of reliable solar wind models for considering the influence of solar wind structures on the event-to-event variability of SEP events. 
In addition, we note that CMEs in the EUHFORIA model are inserted at its inner boundary, located at 0.1~au. Thus, the strong deformation of the CME shock in our modelling is solely due to interactions between the CME and the solar wind beyond 0.1~au. 
However, shock waves can already become distorted in the corona below 0.1~au \citep[e.g.,][]{kwon13}. Such distortions are also expected to significantly alter the efficiency of \RevOne{particle} acceleration along the shock front in the corona, below the starting height of the simulations presented in this work. \citet{Jebaraj22} studied the low-coronal evolution of the same event and suggested that strong EUV wave deformations occurred due to the presence of several magnetic and density structures. 
Taking such deformations into account might help to improve the correlations found by \citet{kouloumvakos19} and \citet{dresing22} between the intensities of observed SEP events and certain shock properties, such as the shock obliquity, which the authors derived from fitting an ellipsoidal shock model to white-light coronagraph images.   
\RevOne{However, it is worth noting that our PARADISE simulation agrees well with the observations, without including an explicit relationship between the shock obliquity and the energetic particle emission from the shock (see Eq.~(\ref{eq:Q})). 
This despite that many particle acceleration mechanisms depend on the shock obliquity \citep[see e.g.,][]{bell78,ball01,chen22}. 
One possible explanation for the accuracy of our simulation is that small scale solar wind turbulence, which EUHFORIA does not account for, causes the shock obliquity to fluctuate significantly as the shock propagates through the solar wind \citep[e.g.,][]{richardson10}.
}
\RevThree{It is also worth mentioning that the SIR and CME shock interaction, which resulted in the initial peak of particle intensity profiles at Bepi and ACE, produced a predominantly quasi-perpendicular shock geometry throughout the event that could have created favourable conditions for efficient acceleration, if a rich suprathermal seed population was present and a small spatial diffusion coefficient perpendicular to the magnetic field increased the acceleration rate of particles \citep[e.g.,][]{jokipii87, giacalone05b, giacalone05c, chen22}.
}

Finally, we note that a good match between the SEP observations and simulations at Bepi was obtained after accounting for the 4-hour mismatch between the observed and simulated CME arrival times.
The importance of bringing the simulated solar wind into agreement with observed solar wind was also pointed out in \citet{wijsen21}, where the authors modelled energetic particle enhancements produced by a SIR that was observed by both Parker Solar Probe (located at 0.56~au) and STEREO-A (near 1~au). 
As in the work presented here, a good agreement between the observed and simulated  energetic particle intensities was only obtained once a mismatch between the modelled and simulated solar wind was taken into account. 
Thus, an important conclusion from these studies is that simulated energetic particle intensities typically tend to show a good agreement with observations only if the underlying solar wind is well captured by the modelling.  
This means that the reliability of any SEP forecasting tool that requires a model for the background solar wind is strongly dependent on the forecasting tool utilised for the underlying ambient medium. 

\section*{Acknowledgements}
N.W.\ acknowledges acknowledges support from NASA program NNH17ZDA001N-LWS and from the Research Foundation - Flanders (FWO-Vlaanderen, fellowship no.\ 1184319N).
D.L.\ and I.G.R.\ acknowledge support from NASA Living
With a Star (LWS) programs NNH17ZDA001N-LWS and
NNH19ZDA001N-LWS, the Goddard Space Flight Center
Internal Scientist Funding Model (competitive work package)
program and the Heliophysics Innovation Fund (HIF) program. 
IGR also acknowledges support from NASA/HSR program NNH19ZDA001NHSR and the ACE mission.
B.S.-C.\ acknowledges support through UK-STFC Ernest Rutherford Fellowship ST/V004115/1 and STFC grant ST/V000209/1.
I.C.J.\ acknowledges funding from the BRAIN-be project SWiM (Solar Wind Modelling with EUHFORIA for the new heliospheric missions) and the European Union's Horizon 2020 research and innovation program under grant agreement No.~870405 (EUHFORIA 2.0).
A.K.\ acknowledges financial support from NASA's NNN06AA01C (SO-SIS Phase-E) contract.
E.P.\ acknowledges support from NASA's Operations to Research (O2R; grant no.\ 80NSSC20K0285) and Living With a Star Strategic Capabilities (LWS-SC; grant no.\ 80NSSC22K0893) programmes.
A.Ar. \ acknowledges the support by the Spanish  Ministerio de Ciencia e Innovaci\'{o}n (MICINN) under grant PID2019-105510GB-C31 and through the ``Center of Excellence Mar\'{i}a de Maeztu 2020-2023'' award to the ICCUB (CEX2019-000918-M)
D.P. acknowledges support by he German Space Agency (Deutsches Zentrum für Luft- und Raumfahrt, e.V., (DLR)) under grant number 50OT2002.
S.P.\ acknowledges support from the projects
C14/19/089  (C1 project Internal Funds KU Leuven), G.0D07.19N  (FWO-Vlaanderen), SIDC Data Exploitation (ESA Prodex-12), and Belspo project B2/191/P1/SWiM.
These results were  also obtained in the framework of the ESA project ``Heliospheric modelling techniques'' (Contract No.\ 4000133080/20/NL/CRS). 
Computational resources and services used in this work were provided by the VSC (Flemish Supercomputer Centre), funded by the FWO and the Flemish Government-Department EWI.

\end{document}